\pdfoutput=1
\documentclass[useAMS,usenatbib]{mn2e}
\usepackage{amsmath, float}
\usepackage{txfonts}
\usepackage{fleqn}   
\usepackage{graphicx}

\newcommand{\be}{\begin{equation}}
\newcommand{\ee}{\end{equation}}
\newcommand{\bdm}{\begin{displaymath}}
\newcommand{\edm}{\end{displaymath}}

\newcommand{\bea}{\begin{eqnarray}}
\newcommand{\eea}{\end{eqnarray}}
\newcommand{\ba}{\begin{align}}
\newcommand{\ea}{\end{align}}
\title[Mass loading in the tails of  BSPWNe]
{Numerical simulations of mass loading in the tails of  Bow Shock Pulsar Wind Nebulae}
\author[B. Olmi, N. Bucciantini, G. Morlino]{
 B. Olmi$^{1,2,3}$ \thanks{E-mail: barbara.olmi@unifi.it}, N. Bucciantini$^{2,1,3}$, G. Morlino$^{4,2}$ 
\\
$^{1}$Dipartimento di Fisica e Astronomia, Universit\`a degli Studi di Firenze, Via G. Sansone 1, 
I-50019 Sesto F.~no  (Firenze), Italy\\
$^{2}$INAF - Osservatorio Astrofisico di Arcetri, Largo E. Fermi 5,
I-50125 Firenze, Italy\\
$^{3}$INFN - Sezione di Firenze, Via G. Sansone 1, I-50019 Sesto F.~no  (Firenze), Italy\\
$^{4}$Gran Sasso Science Institute, Viale Francesco Crispi 7, 67100, L'Aquila, Italy}

\begin{document}
 
\date{Accepted / Received}

\maketitle

\label{firstpage}

\begin{abstract}
When a pulsar is moving through a partially ionized  medium, a fraction of neutral Hydrogen atoms penetrate inside the pulsar wind and can be photo-ionized by the nebula UV radiation. The resulting protons remains attached to the magnetic field of the light leptonic pulsar wind enhancing its inertia and changing the flow dynamics of the wind.  We present here the first numerical simulations of such effect in the tails of bow shock nebulae. We produce a set of different models representative of pulsars moving in the interstellar medium with different velocities, from highly subsonic to supersonic, by means of 2D hydrodynamic relativistic simulations.  We compare the different tail morphologies with results from theoretical models of mass loading in bow shocks. As predicted by analytical models we observe a fast sideways expansion of the tail with the formation of secondary shocks in the ISM. This effect could be at the origin of the {\it head-and-shoulder} morphology observed in many BSPWNe.
\end{abstract}


\section{Introduction}
\label{sec:intro}

Bow shock pulsar wind nebulae (BSPWNe) are a particular type of pulsar wind nebulae (PWNe), usually associated with old pulsars that have emerged from their progenitor supernova remnant \citep{Bucciantini:2001,Gaensler_Slane06a}.  
It has been estimated that a considerable fraction of all the pulsars (from $10\%$ up to $50\%$) is born with kick velocity of the order of 100-500 km s$^{-1}$ \citep{Arzoumanian:2002}. Since the expansion of the surrounding remnant is decelerated due to the sweeping up of interstellar medium (ISM), these pulsars are fated to escape their progenitor shell over typical timescales of few tens of thousands of years, sufficiently short if compared with typical pulsar ages ($\sim 10^6$ years).  After escaping from the remnant those pulsars interact directly with the ISM, and since the typical sound speed in the ISM is of the order of $10-100$  km s$^{-1}$, their motion is highly supersonic.

Pulsars are known to be powerful sources of cold, magnetized and ultra-relativistic outflows, with bulk Lorentz factors in the range $10^4-10^7$ \citep{Goldreich:1969,Kennel:1984,Kennel:1984a}. This wind, mainly composed by electron-positron pairs, is launched from the pulsar magnetosphere \citep{Contopoulos_Kazanas+99a,Spitkovsky06a,Tchekhovskoy_Philippov+16a,Hibschman:2001,Takata_Wang+10a,Timokhin_Arons13a,Takata_Ng+16a}, at the expenses of the rotational energy of the pulsar itself.  The wind, confined and slowed down in a strong termination shock (TS), by the interaction  with the ambient medium gives rise to a nebula,  that may be revealed as a source of non-thermal radiation produced by the accelerated pairs interacting with the nebular magnetic field.

Due to the pulsar fast speed, the ram pressure
balance between the pulsar wind and the supersonic ISM flow (in the
reference frame of the pulsar itself) gives rise to a cometary nebula \citep{Wilkin:1996,Bucciantini:2001,Bucciantini:2002}: the pulsar wind is still slowed down by a TS, elongated behind the pulsar,  beyond which the shocked wind gives rise to the PWN itself, separated from the ISM, shocked in a forward bow-shock, by a contact discontinuity (CD). 
A schematic sketch of this structure can be seen in Fig.~\ref{fig:sketch}. This morphology has been confirmed by numerical simulations in various regimes \citep{Bucciantini:2002,Bucciantini_Amato+05a,Vigelius:2007,Barkov:2018}.

These nebulae can be seen in non-thermal radio/X-rays, usually in the form of  long tails \citep{Arzoumanian_Cordes+04a,Kargaltsev:2017,Kargaltsev_Misanovic+08a,Gaensler_van-der-Swaluw+04a,Yusef-Zadeh_Gaensler05a,Li_Lu+05a,Gaensler05a,Chatterjee_Gaensler+05a,Ng_Camilo+09a,Hales_Gaensler+09a,Ng_Gaensler+10a,De-Luca_Marelli+11a,Marelli_De-Luca+13a,Jakobsen_Tomsick+14a,Misanovic_Pavlov+08a,Posselt_Pavlov+17a,Klingler_Rangelov+16a,Ng_Bucciantini+12a}. In some cases polarimetric informations are also available, suggesting a large variety of magnetic configurations, from highly ordered to strong turbulent, both in the head and in the tail of the nebula \citep{Ng_Bucciantini+12a, Yusef-Zadeh_Gaensler05a, Ng_Gaensler+10a}.

If the pulsar is moving through a partially ionized medium the bow shock may also be detected as H$_\alpha$ emission produced as a consequence of collisional and/or charge exchange excitations of neutral hydrogen atoms in the tail \citep{Chevalier:1980, Kulkarni_Hester88a,Cordes_Romani+93a,Bell_Bailes+95a,van-Kerkwijk_Kulkarni01a,Jones_Stappers+02a,Brownsberger:2014,Romani_Slane+17a}. Such a phenomenon is not restricted to BSPWNe but is observed also in supernova remnant shocks \citep[see, e.g.,][]{Heng:2010,Blasi-Morlino:2012,Morlino-Blasi:2014}.
Recently it was also detected the second example of a BSPWN emitting in the far-UV, with the emission spatially coincident with the previous detected H$_\alpha$ one \citep{Rangelov:2017}.

In the last several years, detailed observations at different wavelengths have revealed a varieties of morphologies in the pulsar vicinity, different emission patterns and shapes of tails, with puzzling outflows misaligned with the pulsar velocity.
This diversity is still poorly understood, and generically attributed to a configuration-dependent growth of shear instabilities and/or to the propagation of the pulsar in a non-uniform ISM \citep{Romani:1997,Vigelius:2007,Yoon-Heinz:2017, Toropina:2018}.

BSPWNe are in particular shown to have a peculiar \textit{head-and-shoulder} structure (shown in the sketch of Fig.~\ref{fig:sketch}), with the smooth bow shock in the head evolving neither in a cylindrical shape nor  in a conical one, rather revealing a sideways expansion that sometimes show a periodic structure (one of the most famous example is the Guitar nebula, see, e.g., \citealt{Chatterjee:2002,Chatterjee:2004,van-Kerkwijk:2008}). While at present the structures in the vicinity of the bow shock head are reasonably well understood, the same cannot be said of the structures in the tail \citep{Brownsberger:2014,Pavan:2016,Klingler:2016}.

The possibility that the dynamics of PWNe can be strongly affected by the mass loading was discussed by \citet{Bucciantini:2001} and \citet{Bucciantini:2002a}. There the authors have shown that a non negligible fraction of neutral atoms is expected to penetrate in the bow shock through shocked ISM, and then interact with the pulsar wind, modifying the predicted dynamics.
These first models provide a good description of both the hydrogen penetration length scale and the H$\alpha$ luminosity. In \citet{Morlino:2015} a quasi-1D steady-state model have been proposed to investigate the effect of mass loading on the tail morphology of BSPWNe. If a significative fraction of neutral atoms from the ISM penetrate into the  bow shock, they can undergo ionization essentially by interacting with the UV photons emitted by the relativistic plasma of the nebula. 
The resulting protons and electrons interact, in turn, with the wind thorough its magnetic field, leading to a net mass loading of the shocked wind in the tail. 
The newly formed protons and electrons are at rest in the rest frame of the ISM. Moreover they originate from the cold neutrals of the ISM which carry negligible thermal energy. Hence they do not add momentum or internal energy  to the flow (they however add their rest mass energy). Given that momentum and energy are conserved, the rise in rest mass energy density causes  the wind to slow down and the pressure to rise, resulting ultimately in a lateral expansion. 
This effect indeed produces a shape which resembles the \textit{head-and-shoulder} morphology. 
Nevertheless the analytic model by \citet{Morlino:2015} has few shortcomings: being 1D it does not account for the fluid dynamics in the lateral direction, the ISM ram pressure is neglected despite the supersonic motion, and, being stationary, it cannot capture possible instabilities.

In this paper we extend this model by means of 2D relativistic HD axisymmetric simulations, with the aim of validating the \textit{head-shoulder} morphology of BSPWNe predicted by the aforementioned work, and extending it beyond its limitations. Although magnetic field might influence the dynamics of the flow,  its response to mass loading, and the expected opening of the tail, in order to avoid a further level of complexity we limit this work to pure HD.

%%%%%%%%%%%%%%%%%%%
%%%% Paper organization %%%%%
%
This paper is organized as follows: in Sec.~\ref{sec:Nsetup} the numerical tool and setup are described; in Sec.~\ref{sec:results} we present and discuss our findings. Conclusions are then drawn in Sec.~\ref{sec:conclusion}.
\section{Numerical setup}
\label{sec:Nsetup}
The typical length scale of BSPWNe, is the so called {\it stand off distance} $d_0$,  \citep{Wilkin:1996,Bucciantini:2001, van-der-Swaluw:2003, Bucciantini_Amato+05a}. This is the distance from the pulsar of the stagnation point where the wind momentum flux and the ISM ram pressure balance each other
\begin{equation}\label{eq:stagnationp}
	d_0 = \sqrt{\dot{E}/(4\pi c \rho_\mathrm{ISM} v_\mathrm{ISM}^2)}\,,
\end{equation}
where $\dot{E}$ is the pulsar luminosity, $\rho_\mathrm{ISM}$ is the ISM density (the ionized component), $ v_\mathrm{ISM}$ the speed of the pulsar with respect to the local medium, and $c$ the speed of light. It is then convenient to normalize all the distances in terms of $d_0$.

\begin{figure}
	\centering
	\includegraphics[width=.45\textwidth]{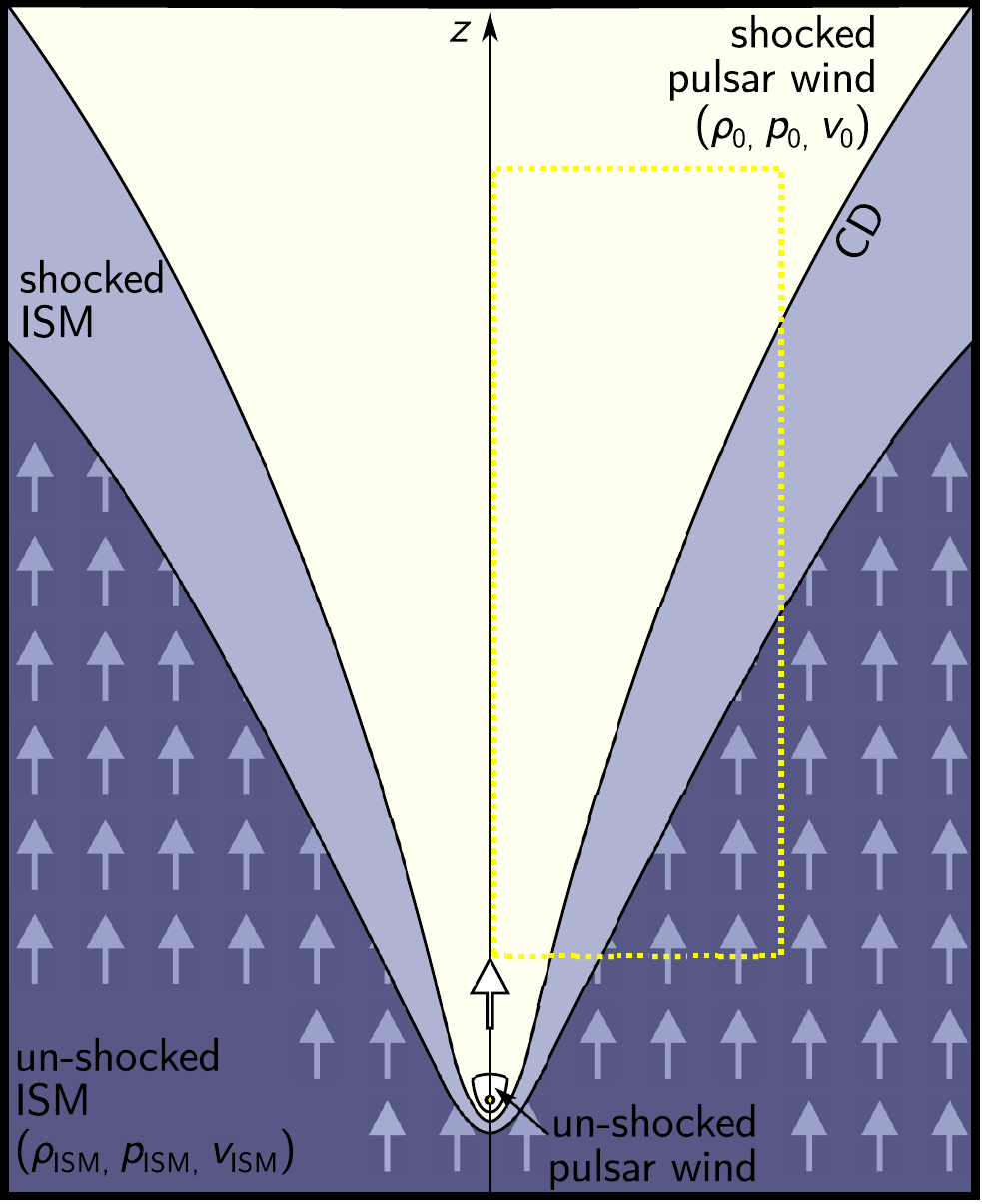}
	\caption{Sketch of the \textit{head-shoulder} structure of a BSN. 
	The simulated region is highlighted with the yellow-dotted rectangle.}
	\label{fig:sketch}
\end{figure}

Numerical simulations \citep{Bucciantini:2002b,Bucciantini_Amato+05a,Bucciantini:2018}, in the absence of mass loading,  show that at a distance of a few $d_0$ from the pulsar a backward collimated tail is already formed.  Typical flow velocities of the shocked pulsar wind plasma in the tail are of the order of $v_0\simeq 0.8c$, while its pressure is approximately in equilibrium with the ISM. Since we are actually interested in the behavior of the tail we do not simulate the bow shock head. In practice we only study the domain shown by the highlighted yellow-dotted rectangle in Fig.~\ref{fig:sketch}.

Simulations were carried out using the numerical code PLUTO \citep{Mignone:2007}, a shock-capturing, finite-volume code for the solution of hyperbolic/parabolic systems of partial differential equations, using a second order Runge-Kutta time integrator and an HLLC Riemann solver. Given the geometry of the problem, simulations were done on  a 2D axisymmetric grid with cylindrical coordinates $(r,\, z)$, assuming a vanishing azimuthal velocity $v_\phi=0$.  The grid has uniform spacing along $r$ and $z$, but with different resolution along the two directions, in order to ensure the requested numerical resolution in the radial direction at the injection zone.

The fluid is described using an ideal gas equation of state (EoS) with adiabatic index $\Gamma=4/3$, appropriate for the relativistic shocked pulsar plasma.  \citet{Bucciantini:2002} has shown that more sophisticated simulations with multi-fluid treatment, enabling the use of a different adiabatic coefficient for the non-relativistic component, lead only to minor deviations in the overall geometry, until efficient thermalization between the two components is reached, which physically happens on timescales much longer than the flow time in the tail of BSPWNe. For this same reason the use of {\it adaptive EoS}, like Taub's EoS \citep{Taub:1948, Mignone:2005}  is not feasible in the presence of mass loading (or other form of mass contamination), because they are based on the assumption of instantaneous thermalization between the two components.

Our reference frame is moving with the pulsar. In this reference frame the ISM is seen as a uniform, unmagnetized flow moving along the positive $z-$direction with velocity $v_z=v_\mathrm{ISM}$. The unloaded pulsar tail plasma is also moving along the positive $z-$direction with velocity $v_z=0.8c$. In accord with \citet{Morlino:2015}, the unloaded tail is supposed to extend between $r=0$ and $r=1$ (in units of $d_0$).
As we will show the parameters regulating the tail geometry are the ISM pressure $p_\mathrm{ISM}$ and the ISM Mach number $M^2 = \rho_\mathrm{ISM} v_\mathrm{ISM}^2/(\Gamma p_\mathrm{ISM})$, and not the separated values of the ISM density and velocity. At the $z=0$ boundary in a layer extending up to $z=1.5$ we keep fixed the inflow conditions for the ISM and pulsar wind tail, in terms of velocity density and pressure. The pressure in the tail is set equal to the one in the ISM, $p_{\rm tail}=p_\mathrm{ISM}$, while its density is $\rho_{\rm tail}= 10^{-2} p_{\rm tail}$ appropriate for the hot shocked relativistic gas of the PWN (we verified that lowering it further does not change the results). The velocity of the ISM is indeed varied in order to sample both the subsonic and the supersonic regimes with Mach numbers ranging from 0 to $\sim 6$ (see Tab.~\ref{Tab:Runs}). Axial symmetry is imposed in $r=0$, while outflow conditions are set at the other two boundaries.

\begin{table}
\begin{center}
\begin{tabular}{{cccccccc}}%{| l | c | c | c | c | c | p{0.5cm} |}
\hline
 Run  & $(r_f,\, z_f)$ [$d_0$]  & $(N_r,\,N_z)$ & $\rho_\mathrm{ISM}$   & $v_\mathrm{ISM}$ [$c$]  & $M$ \\
\hline
 S$_0$	& 	$(15,\,50)$ 	& 	$(320,\,320)$  	& 1.00         &  $0$		&  $0.0$\\
 S$_1$	& 	$(15,\,60)$ 	& 	$(320,\,320)$ 	& 1.00         &  $0.015$	&  $0.4$\\
 S$_{1a}$	& 	$(15,\,60)$ 	& 	$(320,\,320)$  	& 0.25         &  $0.03$	&  $0.4$\\
 S$_2$	&	$(15,\,60)$ 	& 	$(320,\,320)$  	& 1.00         &  $0.03$	&  $0.8$\\
 S$_{2a}$	&	$(15,\,60)$ 	& 	$(320,\,320)$  	& 0.25         &  $0.06$	&  $0.8$\\
 S$_3$	& 	$(10,\,100)$ 	& 	$(256,\,512)$  	& 1.00         &  $0.07$	&  $1.9$ \\
 S$_4$	& 	$(10,\,100)$ 	& 	$(256,\,512)$  	& 1.00         &  $0.1$ 	&  $2.7$\\
 S$_5$	& 	$(10,\,100)$ 	& 	$(256,\,512)$ 	& 1.00         &  $0.14$ 	&  $3.8$\\
 S$_6$	& 	$(10,\,100)$ 	& 	$(256,\,512)$ 	& 1.00         &  $0.2$ 	&  $5.5$\\
 S$_{6a}$	& 	$(10,\,100)$ 	& 	$(256,\,512)$ 	& 4.00         &  $0.1$ 	&  $5.5$\\
 S$_{6b}$	& 	$(10,\,100)$ 	& 	$(256,\,512)$ 	& 2.04         &  $0.14$	&  $5.5$ \\
\hline
\end{tabular}
\end{center}
\caption{List of all runs. The columns are from the left: box size $r_f$ and $z_f$ in unit of $d_0$; number of grid points $N_r$ and $ N_z$ in the $r$ and $z$ directions; ISM density (arbitrary units) and velocity (in unit of $c$); pulsar Mach number $M$.}
\label{Tab:Runs}
\end{table}
%
%
 %%%%%%%%%%%%%%%%%%%% Fig 1 %%%%%%%%%%%%%%%%%%%%%%%%%
%%%%%%%%%%%%%%%%%%%%%%%%%%%%%%%%%%%%%%%%%%%%%%%%

\subsection{Mass loading}
\label{SS:mload}
Given that mass loading is supposed to act only on the shocked pulsar wind material (as discussed in \citealt{Morlino:2015}), we also evolve a numerical tracer $\chi$ that allows us to distinguish it with respect to the ISM. 
This ensures that we can identify the bow shock tail region even when it starts to broaden.

The tracer has value 1 in the BSPWN tail and 0 for the ISM.
The additional mass is loaded uniformly along the spatial grid, wherever it is satisfied the condition $\chi=1$.

Given that the typical pulsar kick velocities in the ISM are much smaller than the typical flow speed in the tail, 
we simplify the treatment of mass loading, assuming that the this mass is added with zero momentum. This approximation is valid as long as the loaded tail does not slow down to speed comparable to $v_{\rm ISM}$.

The effect of the mass loading can then be simulated with a simple modification of the equation for  mass conservation  (PLUTO evolves the reduced energy density that does not include the rest mass energy density)  according to
\begin{equation}\label{eq:rhos}
	\frac{\partial (\gamma \rho)}{\partial t} + \vec{\nabla} (\gamma \rho \vec{v})= \chi \dot{\rho}\,,
\end{equation}
where $\dot{\rho}$ is the rate of mass loading that depends on the ISM density of neutrals and the UV ionizing flux from the PWN \citep{Morlino:2015}. In the present work we adopt a value of $\dot{\rho}$ such that  the distance $\lambda_{\rm rel}$  over which the inertia in the BSPWN tail doubles with respect to the equilibrium value is
\begin{equation}\label{eq:lambda}
\lambda_\mathrm{rel} = \frac{4p_\mathrm{ISM} v_0}{\dot{\rho}} = 8 d_0\,.
\end{equation}
This value has been chosen in order to optimize the simulation runs in terms of speed and size of the computational domain. Notice also that $\lambda_{\rm rel}$ is assumed to be constant in space and time while, in principle, it can change depending on the  amount of neutral atoms at each spatial position, which is equivalent to the requirement that only a small fraction of the ISM neutral are ionized. 

\section{Results and Discussion}
\label{sec:results}

The mass loading has a twofold effect on the tail dynamics: the density in the tail increases according to Eq.~(\ref{eq:rhos}) and the bulk velocity decreased, since the total momentum is conserved. As a consequence the tail pressure rises, and this leads to an expansion of the tail cross section, until a new steady state is reached. All the results are shown when the stationarity is reached, namely when  no significant evolution is observed anymore. 

\begin{figure}
	\centering
	\includegraphics[width=.5\textwidth]{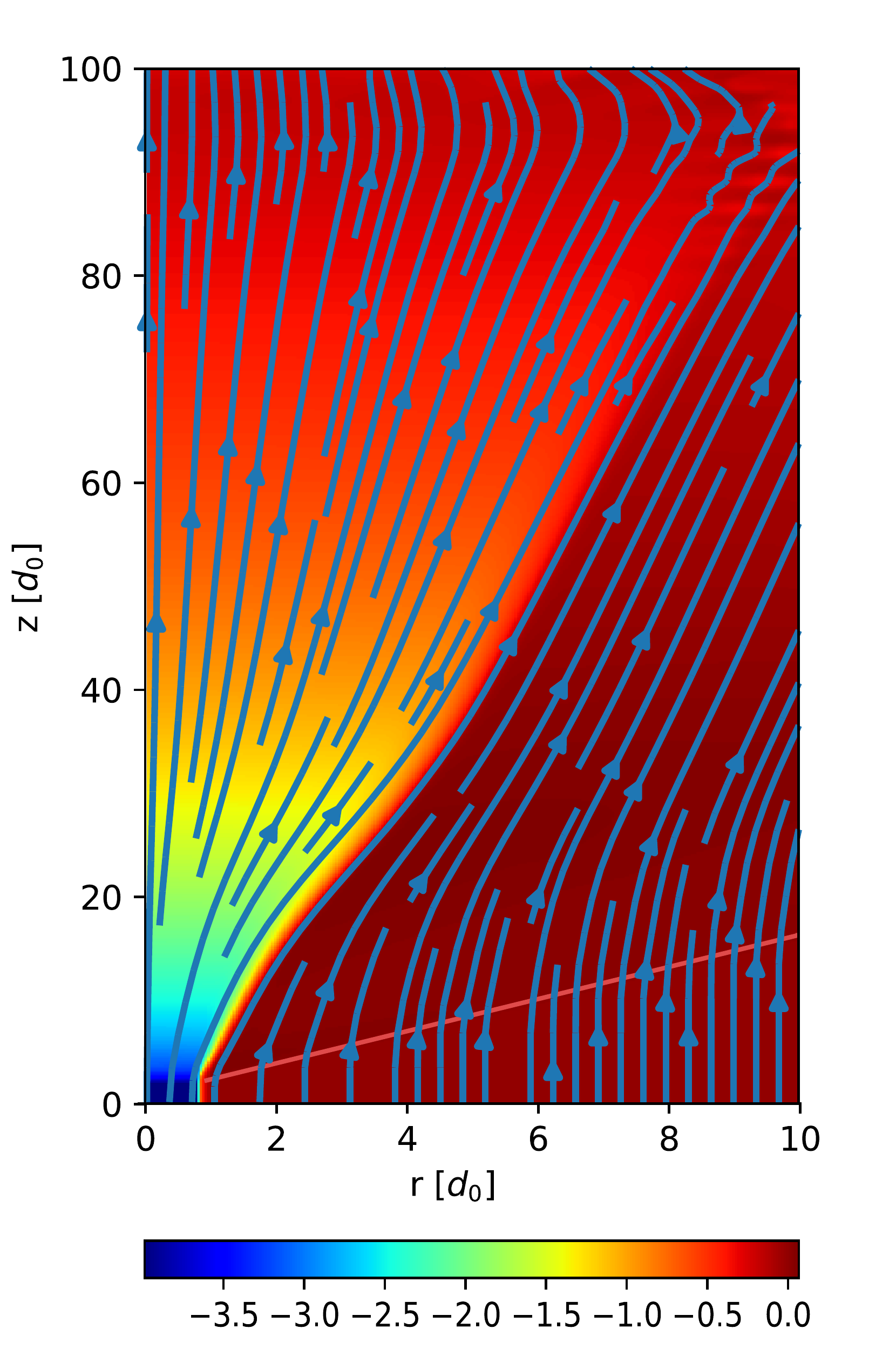}
	\caption{Logarithmic map of the density for run S$_3$, chosen as representative, with velocity streamline superimposed. 
	The effect of the radial component of the velocity on the velocity streamlines is clearly visible.The light-red line identifies the presence of the forward shock in the ISM medium.}
	\label{fig:STRM}
\end{figure}

In Table~\ref{Tab:Runs}, we list all the runs. In Fig.~\ref{fig:STRM} we show the morphology of the tail (density and streamlines) for the representative run S$_3$.  In all supersonic runs, the expansion of the tail is accompanied by the generation of an oblique shock in the ISM, also shown in the same figure. Such oblique shocks are indeed predicted whenever the contact discontinuity makes an angle with the flow that is larger than the Mach cone \citep{Morlino:2015}.

 %%%%%%%%%%%%%%%%%%%% Fig 2 %%%%%%%%%%%%%%%%%%%%%%%%%
\begin{figure}
	\centering
	\includegraphics[width=.5\textwidth]{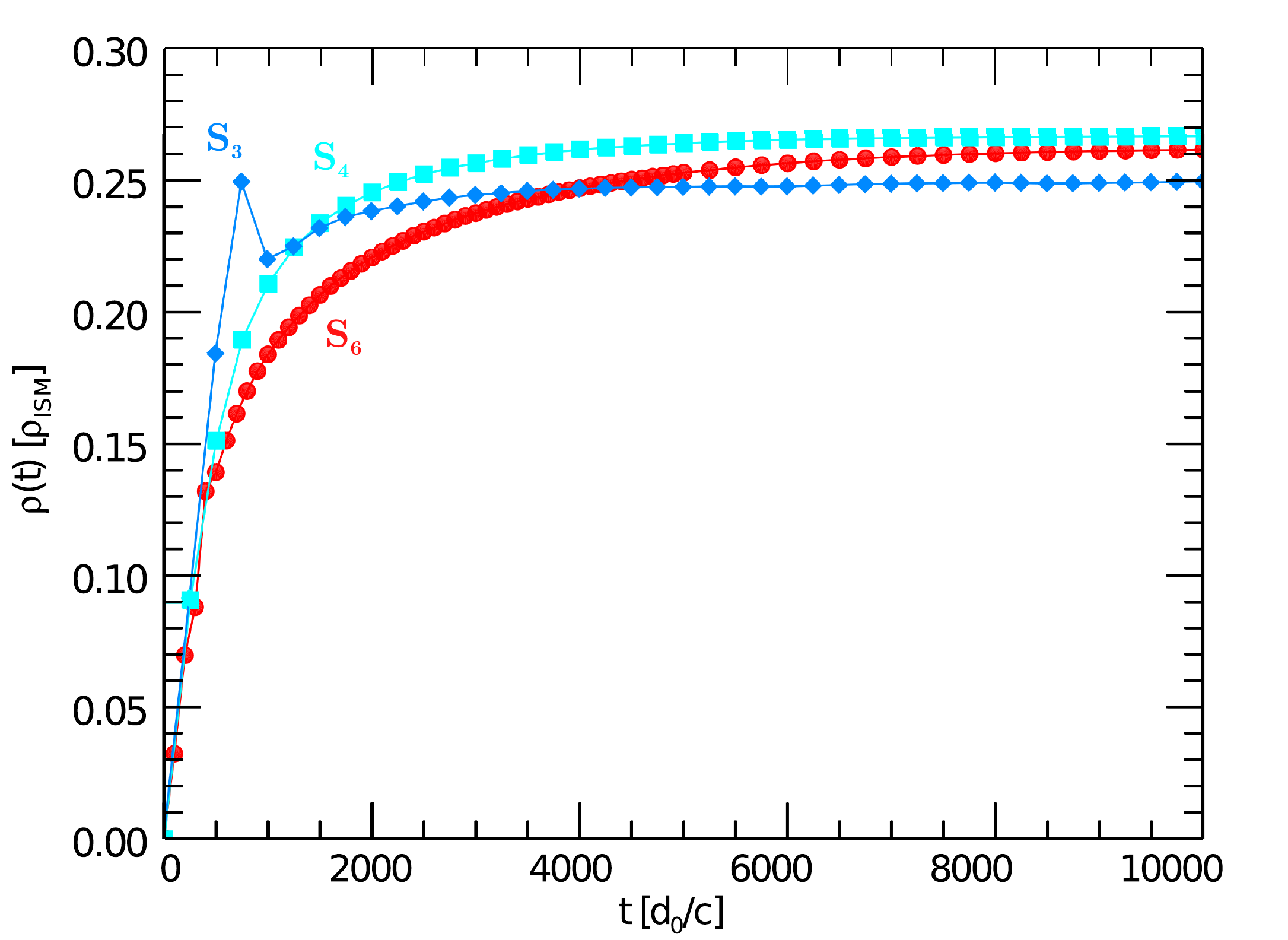}
	\caption{Variation of the mass density with time for the runs S$_3$, S$_4$ and S$_6$, chosen as illustrative cases. The value is computed at $r=0.4$ as an average on a few neighbor cells around $z=60$. Starting from the initial mass of the shocked wind ($\rho_0=10^{-4}\rho_\mathrm{ISM}$) the effect of the mass loading  is clearly visible. As time passes, the density increases up to the saturation value that is a fraction of the ambient medium density $\rho_\mathrm{ISM}$.}
	\label{fig:MvsT}
\end{figure}

%%%%%%%%%%%%%%%%%%%%%%%%%%%%%%%%%%%%%%%%%%%%%%%%
%
In Fig.~\ref{fig:MvsT}, we show how the density, taken at a fixed position in the tail,  changes in time. Starting from the initial value of $\rho_{\rm tail}$, the tail density increases until it reaches the asymptotic value of $\sim \rho_\mathrm{ISM}/4$ (this value depends on the position). We see that the steady state is reached already at $t \simeq 5000 d_0/c$, when the system approaches a stationary configuration in the entire domain. Afterward, the density remains almost constant and the tail preserves its morphology. This is a good estimate of the time it takes to reach steady state for all the cases we run.
%
 %%%%%%%%%%%%%%%%%%%% Fig 3 %%%%%%%%%%%%%%%%%%%%%%%%%
%%%%%%%%%%%%%%%%%%%%%%%%%%%%%%%%%%%%%%%%%%%%%%%%
%

\begin{figure}
	\centering
	\includegraphics[width=.5\textwidth]{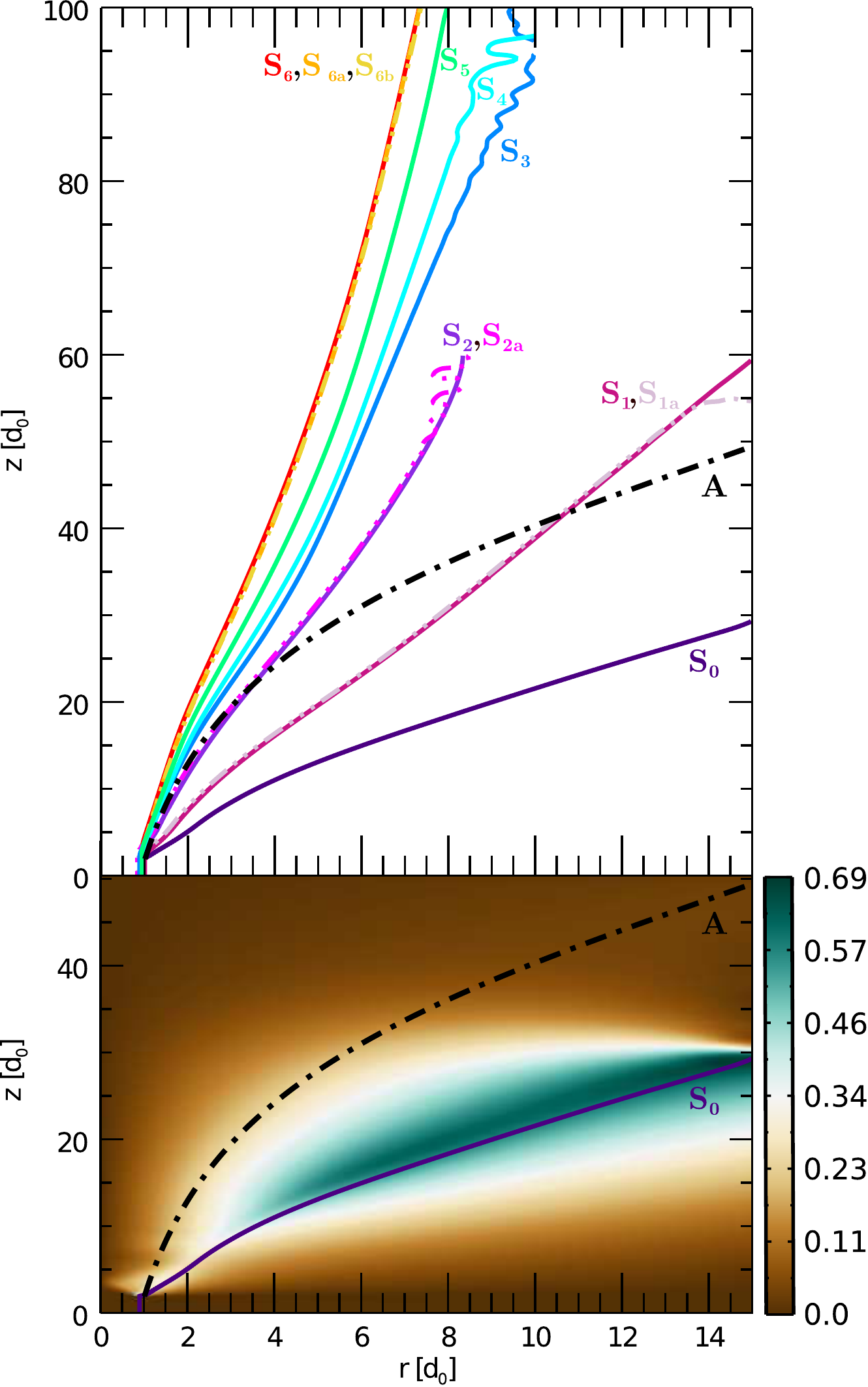}
	\caption{Top panel: density contours for the all the considered configurations S$_i$ listed in Table~\ref{Tab:Runs}. 
	The A curve is the analytical profile obtained from \citep{Morlino:2015}, to be compared with S$_0$ run.
	The $z$ and $r$ axis are plotted in units of the stagnation point $d_0$. Solid lines indicate different runs and thus different values of $v_\mathrm{ISM}$. As expected the tail appears to be more broaden when the velocity of the ambient medium is smaller, namely when the pulsar velocity is subsonic (S$_0$ is the one in the limit $M=0$). On the contrary, as the Mach number grows beyond unity (run S$_2$ has $M\simeq1.9$) and the pulsar enters in the supersonic regime, the tail starts to be more narrow. 
	Bottom panel: the same density profiles for the analytic model and S$_0$ run superimposed to a colored map of $v_r/v_z$. The extra broadening of S$_0$ with respect to A can be seen as the effect of the radial component of the velocity, and its related momentum, which is not present in the analytical model.}
	\label{fig:tailsCFR}
\end{figure}

In the top panel of Fig.~\ref{fig:tailsCFR} the position of the contact discontinuity, once the steady state is reached,  is shown for all the configurations listed in Table \ref{Tab:Runs}. For each simulation the contact discontinuity is identified as contour lines of the inertia ($\rho v^2$). Different runs are drawn as lines of different colors and labelled as S$_i$, to be compared with Table \ref{Tab:Runs}, for an easier identification of the current configuration. 
%
 %%%%%%%%%%%%%%%%%%%% Fig  4 %%%%%%%%%%%%%%%%%%%%%%%%%
%%%%%%%%%%%%%%%%%%%%%%%%%%%%%%%%%%%%%%%%%%%%%%%%

As a first step we begin by comparing our results with the analytical model by \citet{Morlino:2015}.
Notice that in principle the velocity of the loaded material is always equal to the ISM velocity, and this is assumed in the cited work, while in our simulations the extra mass is loaded with velocity equal to zero. However it is easy to show that, in the limit of zero speed, the model by  \citet{Morlino:2015}, admit a simple analytical solution which is
\begin{flalign} 
    v(z)         &= v_0 \operatorname{e}^{-z/\lambda_\mathrm{rel}}\,,  && \label{eq:an_gio_v} \\ 
    \rho_p(z) &= \left(\frac{4 p_\mathrm{ISM}}{c^2}\right) 
    			\left( \operatorname{e}^{z/\lambda_\mathrm{rel}}-1\right)\,,  &&  \label{eq:an_gio_rho}  \\
    R(z)        &= R_0 \operatorname{e}^{z/(2\lambda_\mathrm{rel})} \,,  && \label{eq:an_gio_R} 
\end{flalign}
where $R_0=d_0$ is the position of the contact discontinuity at the initial time. This solution agrees to the more general case as long as the flow in the tail remains higher than the ISM speed. In our simulations we find that this conditions holds in our entire domain. The function $R(z)$ is shown in the top panel of Fig.\ref{fig:tailsCFR}, to
be compared with our numerical result for $v_\mathrm{ISM}=0$.  
 This should be taken as a limit for small pulsar speeds, given that for $v_\mathrm{ISM}=0$ no tail is expected in a realistic situation. Moreover this condition corresponds to the assumption, adopted in the analytic model by \cite{Morlino:2015}, that the ISM ram pressure is negligible at the interface between the pulsar wind material and the ISM, allowing us a direct comparison and check of those results.
One can clearly see that the broadening of the two curves is very similar, but the analytic model starts to broaden at a slightly larger distance, from the injection point. This difference is likely due to the 1D assumption of the analytical model that neglects the transverse component of the velocity. From the bottom panel of Fig.~\ref{fig:tailsCFR} one can clearly see that ratio between the transverse (radial) and aligned ($z$) components of the velocity, $v_r/v_z$, reaches values of the order of 0.5. To this transverse velocity there is an associated transverse ram pressure, that is comparable to the bounding ISM pressure. This term, neglected in the work by \citet{Morlino:2015}, exert a pressures that facilitates the opening of the system. 

If we consider the behavior of the analytical solution $A$ and the numerical one S$_0$ in Fig.~\ref{fig:tailsCFR}, with reference to the ratio $v_r/v_z$, we can see that the analytic curve $A$ lies at the boundary of the region in which the $v_r$ component becomes comparable to $v_z$.

\begin{figure}
	\centering
	\includegraphics[width=.45\textwidth]{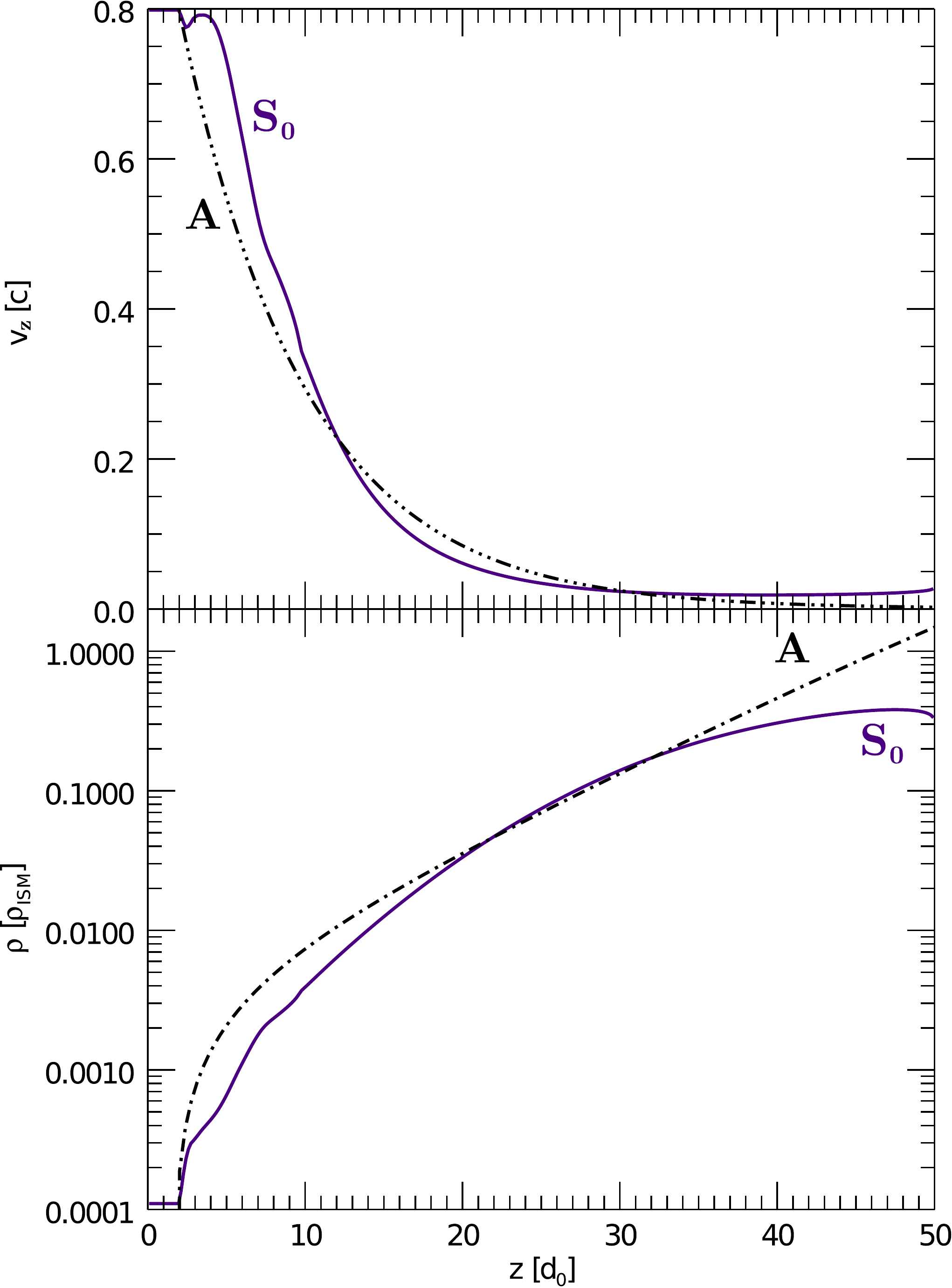}
	\caption{Comparison between the analytical model by \citet{Morlino:2015} (black dot-dashed lines) and simulation S$_0$ (solid-indigo line). In the top panel the $z$ component of the velocity, in $c$ units, is given as a function of $z$, in $d_0$ units. In the bottom panel the logarithmic plot of the density (in $\rho_\mathrm{ISM}$ units) is shown.}
	\label{fig:cfrAS0}
\end{figure}

In Fig.~\ref{fig:cfrAS0} the analytical profiles for $v_z$ and the density, as given by Eqs.~(\ref{eq:an_gio_v})-(\ref{eq:an_gio_rho}), are compared with those from case S$_0$, taken along the axis of the tail.  
As can be seen in Fig.~\ref{fig:STRM} the fluid quantities, like pressure and density, stay almost uniform across the tail at fixed $z$, with the obvious exception of the region in the vicinity of the contact discontinuity. 
One can clearly see that the agreement with the analytical predictions is quite good except close to the outer boundary (probably because of boundary effects) and close to the injection region, where the condition in the tail changes rapidly as a consequence of mass loading.

 %%%%%%%%%%%%%%%%%%%% Fig 4 %%%%%%%%%%%%%%%%%%%%%%%%%
%%%%%%%%%%%%%%%%%%%%%%%%%%%%%%%%%%%%%%%%%%%%%%%%

%---------------------------------------------------------

Actually pulsars move in general through the ISM with supersonic speeds.
This means that the ISM ram pressure can not be taken as negligible with respect to the thermal pressure in confining the tail. In Fig.~\ref{fig:tailsCFR}, we show the result of various run with increasing Mach numbers, up to the value $M \sim 6$, including a case 
close to the transonic regime (S$_{2a}$). As can be seen, the shape of the tail (the opening of the CD) is just a function of the Mach number and not of the ISM density and velocity. This ensures that our runs, with values of the ISM velocity much in excess of realistic numbers, produce correct outcomes, and that it is possible to speed up the computation without compromising the results. This moreover confirms that not only the shape of the head of the bow-shock is just a function of the  Mach number \citep{Wilkin:1996} but also the tail. 
We cannot however rule out that the speed difference between the tail and the ISM could play a key role in the development of shear instabilities, which are only marginally observed in our runs, as testified by the oscillating pattern of the CD close to the outer boundary.

Looking at the morphology of the tail in the different cases we can see that, as the Mach number increases, the tail became less and less broaden. In particular going from $M=0$ to $M=1$ we see a change in the opening angle of about a factor 2, while rising further the mach number to 6 does not provide a proportionally larger collimation. 
It appears that the effect of the ISM ram pressure becomes less and less effective for high $M$.  
Subsonic cases are also more affected by turbulence at the contact discontinuity, that tends to destroy the tail at large distances from the bow shock head. For this reason we selected our simulation boxes for the subsonic cases in order to keep the tail coherent. The same effect is indeed much less evident for the supersonic cases, in which the tail remains coherent up to larger distances from the origin.

\section{Conclusions}
\label{sec:conclusion}
In this paper for the first time we simulated the tail of bow shock pulsar wind nebulae in presence of mass loading due to neutrals in the external medium. This is bound to happen whenever a neutron star moves through a partially neutral medium, where the interaction of the neutral component of the ISM with the relativistic pulsar wind plasma results in a mass loading of the latter, modifying its dynamical evolution. This effect can be important to explain the large variety of shapes observed in the BSPWNe, especially concerning the {\it head-and-shoulder} shape observed in some of them.

The effect of mass loading is studied here by means of 2D relativistic HD simulations using the PLUTO code. A dedicated module that handle the mass loading in relativistic regime has been developed and added to the original code. We explored several configuration for pulsars moving both subsonically and supersonically through the ISM and we assume that the new mass is loaded with null initial velocity. In all case we found that when the inertia of the loaded mass starts to dominate over the inertia of the relativistic wind, the wind slows down and undergoes a sideway expansion, confirming previous findings by analytical models. However, we have shown that the ram pressure of the ISM, neglected in analytical models, can instead strongly modify the expansion properties, and this is particularly relevant in the highly supersonic case, that characterizes known pulsars. Interestingly, the expansion of the tail is only a function of the pulsar Mach number with respect to the ISM and decreases with increasing Mach number as a consequence of the increasing ram pressure exerted by the ISM on the contact discontinuity between the wind and the shocked ISM. For the supersonic cases we also observe the development of oblique shocks in the ISM due to the tail expansion. It might be feasible to add this effect to the analytical model, and we plan to do it in the future. 
Our results however show that it is unlikely that a large radial expansion of the tail (exceeding a few) could be achieved for realistic pulsar-ISM conditions, given that pulsars move in general with high Mach number through the ISM.
On the other hand, our simulations extend up to a distance where the velocity of the wind is $\sim 0.02 c$, a value still much larger than the typical pulsar speed and such that the shear between the pulsar wind and the ISM is still strongly supersonic. In such a conditions instabilities at the contact discontinuity do not grow efficiently \citep{Bodo-Mignone+:2004}. It is than possible that when the flow speed keeps slowing down, reaching a value of the order of $v_{\rm ISM}$ and the relative shear becomes subsonic, Kelvin-Helmholtz instabilities start growing efficiently disrupting the laminar dynamics in the tail and leading to the formation of large bubbles.

In the present work we have neglected the role of the magnetic field, which, depending on its initial configuration, could become dynamically important when the plasma is compressed, as we observe in the tail.  If the flow in the tail is assumed to be laminar at injection and to remain so, then 2D simulations could still be used to investigate the magnetic field dynamics, even if the parameter space can substantially increase given that the strength and geometry of the field now enters as new parameters. However, it is well known that magnetic field tend to make the flow more unstable \citep{Mizuno:2011,Mignone:2013}, and in this case the dynamics can only be handled with full 3D simulations.

\section*{Acknowledgements}

We acknowledge the author of the PLUTO code \citep{Mignone:2007}  that  has  been  used  in  this  work  for  HD simulations. The authors also acknowledge support from the PRIN-MIUR project prot. 2015L5EE2Y "Multi-scale simulations of high-energy astrophysical plasmas".

\footnotesize{
\bibliographystyle{mn2e}
\bibliography{olmi}
}

\end{document}